\documentclass[a4paper,11pt]{article}

\usepackage{pos}
\usepackage{braket} 
\usepackage{subfig} 
\usepackage{array,booktabs,ragged2e} 
\usepackage{xspace} 

\def\amupole{\ensuremath{a_\mu^{p\mathrm{-pole}}}\xspace}
\def\amupipole{\ensuremath{a_\mu^{\pi^0\mathrm{-pole}}}\xspace}
\def\amuetapole{\ensuremath{a_\mu^{\eta\mathrm{-pole}}}\xspace}
\def\amuetappole{\ensuremath{a_\mu^{\eta^{\prime}\mathrm{-pole}}}\xspace}
\def\etap{\ensuremath{\eta^{\prime}}\xspace}

\title{Pseudoscalar transition form factors and the hadronic light-by-light contribution to the muon $g-2$}
\ShortTitle{Pseudoscalar TFFs and the HLbL contribution to the muon $g-2$}

\manuallySeparateAuthors
\author[a]{Antoine G\'erardin,}
\author[b]{ Jana N. Guenther,}
\author[b]{ Lukas Varnhorst}
\author{ and}
\author*[a]{ Willem E. A. Verplanke}
\author{ for the Budapest-Marseille-Wuppertal Collaboration}

\affiliation[a]{Aix-Marseille Universit\'e, Universit\'e de Toulon, CNRS, CPT,\\
	Marseille, France}

\affiliation[b]{Department of Physics, University of Wuppertal,\\
	D-42119 Wuppertal, Germany}

\emailAdd{willem.verplanke@cpt.univ-mrs.fr}
\emailAdd{antoine.gerardin@cpt.univ-mrs.fr}

\abstract{We report on our progress toward the computation of the $\pi^0$, $\eta$ and $\etap$ transition form factors using staggered quarks on $N_f=2+1+1$ gauge ensembles generated by the Budapest-Marseille-Wuppertal collaboration. These form factors are essential ingredients to evaluate the pseudoscalar-pole contributions to the hadronic light-by-light scattering in the muon $g-2$. Preliminary results for the pseudoscalar-pole contributions are presented, at finite lattice spacing, for all three light mesons.}

\FullConference{%
The 39th International Symposium on Lattice Field Theory,\\
8th-13th August, 2022,\\
Rheinische Friedrich-Wilhelms-Universität Bonn, Bonn, Germany
}

\begin{document}
\maketitle

\section{Introduction}
\label{sec:intro}
\begin{figure}[]
	\centering
	\includegraphics[width=0.72\textwidth]{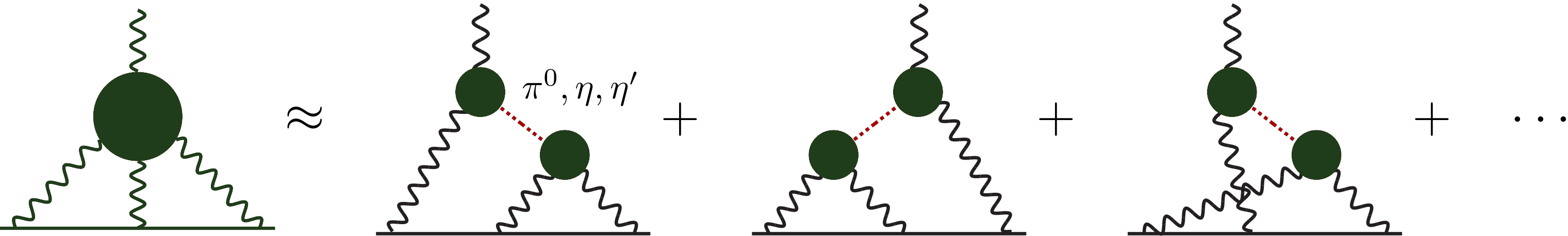}
	\caption{Hadronic light-by-light diagram and its decomposition into the dominant pseudoscalar poles. A wobbly line indicates a photon, the straight line the muon and a blob the non-perturbative hadronic interactions encoded in the pseudoscalar transition form factor.}
	\label{fig:hlbl}
\end{figure}
\noindent The error budget of the anomalous magnetic moment of the muon $a_\mu$ is dominated by two hadronic contributions: the leading order Hadronic Vacuum Polarization (HVP) and the Hadronic Light-by-Light (HLbL) scattering. In spite of the former being $\mathcal{O}(\alpha_e^2)$ and the latter $\mathcal{O}(\alpha_e^3)$, errors are comparable in size. Two model-independent approaches have been proposed to compute the HLbL diagram: the direct lattice computation \cite{Chao:2021tvp, Blum_2020} and the data-driven dispersive approach \cite{Colangelo:2014dfa, Colangelo:2014pva, Colangelo:2015ama}. Crucial input for the dispersive approach are the transition form factors (TFFs) for which relatively little is known from experiment. These TFFs are related to pseudoscalar-pole contributions to $a_\mu^{\mathrm{HLbL}}$ through \cite{Knecht_2002} (see also Figure \ref{fig:hlbl})
\begin{gather}\raisetag{\baselineskip}\begin{aligned}
	\amupole = \left(\frac{\alpha_e}{\pi}\right)^3 \int_{0}^{\infty}dQ_1 \int_{0}^{\infty}dQ_2\int_{-1}^{1}&d\tau \, \left[w_1(Q_1,Q_2,\tau)\mathcal{F}_{p\gamma^*\gamma^*}(-Q_1^2,-Q_3^2)\mathcal{F}_{p\gamma^*\gamma^*}(-Q_2^2,0)\right.\\
 +& \left.w_2(Q_1,Q_2,\tau)\mathcal{F}_{p\gamma^*\gamma^*}(-Q_1^2,-Q_2^2)\mathcal{F}_{p\gamma^*\gamma^*}(-Q_3^2,0)\right],
 \label{eq:amupole}
\end{aligned}\end{gather}
where $Q_3^2 = Q_1^2+Q_2^2 + 2\tau Q_1Q_2$ and $\tau = \cos\theta $ with $\theta$ the angle between $Q_1$ and $Q_2$. $w_i(Q_1,Q_2,\tau)$ are analytically known weight functions peaked at low spacelike $Q^2$. The \amupole receives contributions from three pseudoscalar mesons: the $\pi^0$, $\eta$ and $\eta^{\prime}$. The $\pi^0$-pole contribution has been determined on the lattice in \cite{Gerardin:2016cqj, Gerardin:2019vio}, preliminarily in \cite{Burri:2021cxr} and in the data-driven dispersive framwork \cite{Hoferichter_2018}. The two different methods yield compatible results for this contribution. For the $\eta,\etap$ mesons there is neither a lattice nor a dispersive result yet (though preliminary results have been shown in \cite{burri_muong2,Verplanke:2021gat, verplanke_muong2}). The challenges for lattice QCD regarding these observables are the mixing between the $\eta,\etap$ and noisy, sizable disconnected diagrams. As the weight functions are peaked at low spacelike $Q^2$, information about the TFFs in this regime is crucial for a precise determination of the \amupole contributions.

\subsection{Experimental Data}
\noindent We know that the normalization of the TFF of a pseudoscalar meson is related to a partial decay width $\Gamma(p \to \gamma \gamma)$,
\begin{equation}
	\Gamma(p\to \gamma \gamma ) = \frac{\pi \alpha_e^2 m_p^3}{4}\mathcal{F}_{p \gamma^* \gamma^*}(0,0),
\end{equation}
where $m_p$ is the mass of the meson. Current values are $\Gamma(\pi^0 \to \gamma \gamma) = 7.80(12)$ eV \cite{PrimEx-II:2020jwd}, $\Gamma(\eta \to \gamma \gamma) = 0.516(18)$ keV \cite{ParticleDataGroup:2020ssz},  $\Gamma(\eta^{\prime} \to \gamma \gamma) = 4.28(19)$ keV \cite{ParticleDataGroup:2020ssz}. The two-photon decay widths have been measured with a relative precision of a few percent and can be used to constrain the lattice data. Such a constraint has already been tested in \cite{Gerardin:2019vio} and showed a reduction in the error on \amupipole by more than 30\%. 
\begin{figure}
	\centering
	\includegraphics[width=0.9\textwidth]{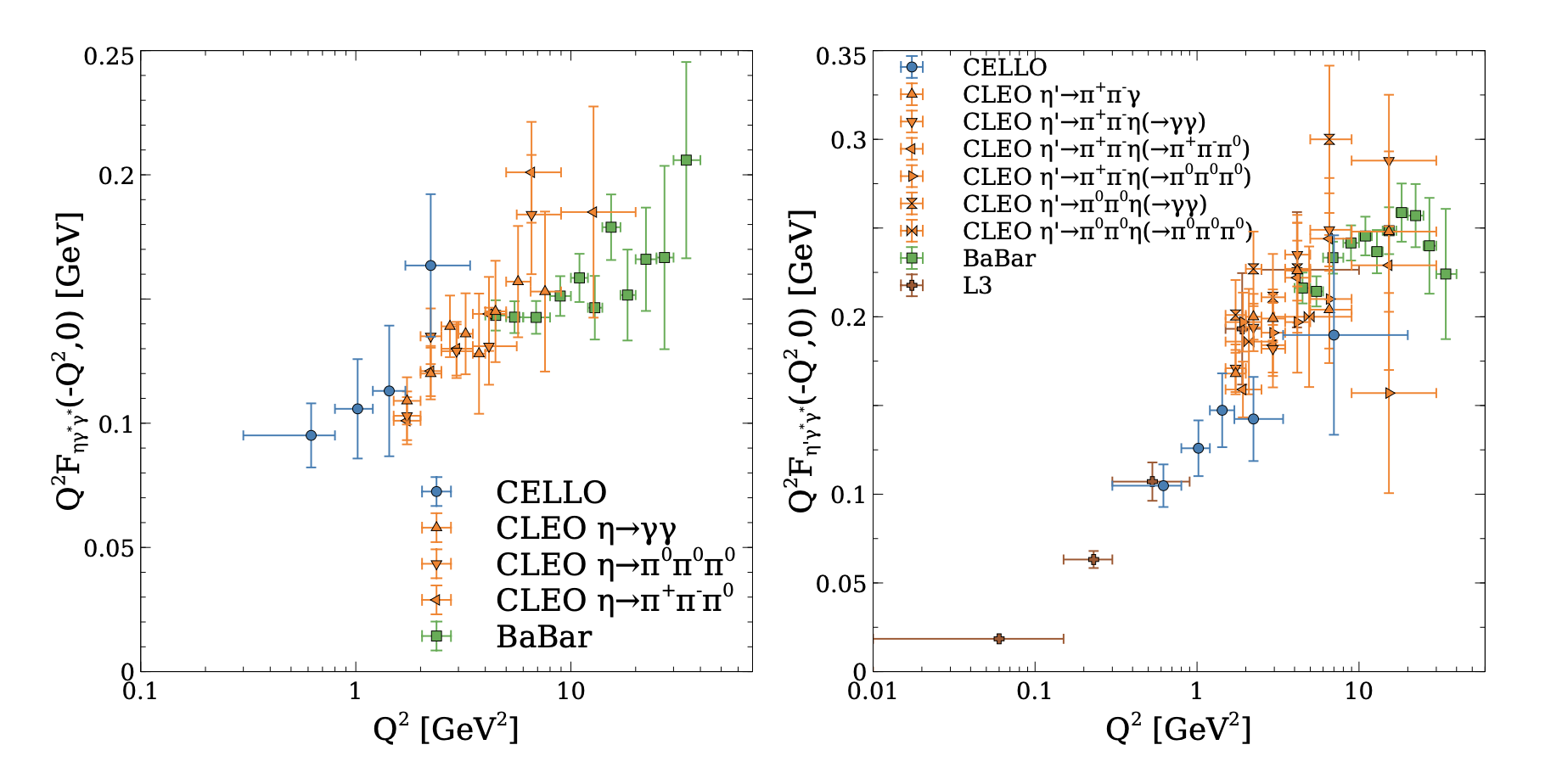}
	\caption{Experimental data on the spacelike TFFs of the $\eta$ (left) and $\eta^{\prime}$ (right) mesons. BABAR \cite{BABAR_exp} (green boxes), CELLO \cite{CELLO:1990klc} (blue circles), CLEO \cite{CLEO_1998} (yellow triangles) and L3 data \cite{L3} (brown crosses) are plotted. Figure extracted from \cite{Aoyama:2020ynm}. }
	\label{fig:exp_data}
\end{figure}

There is also a large amount of experimental data available in the spacelike regime of the TFF for the singly virtual case (one  real photon) as shown in Figure \ref{fig:exp_data}, but at $Q^2\gtrsim 1$ GeV$^2$. When both photons are virtual (doubly virtual), there is no data avalaible below $6$ GeV$^2$. The absence of precise data at low $Q^2$ is crucial, as aforementioned, because this is the important regime for \amupole. Here lattice QCD jumps in, and can provide valuable low $Q^2$ data that is typically challenging to obtain in experiment. Then, a combination of the lattice and experimental data could provide an interesting comparison with the pure lattice result. 
\subsection{Simulation details}
\noindent We use $N_f = 2+1+1$ dynamical staggered fermions with four steps of stout smearing generated by the Budapest-Marseille-Wuppertal collaboration \cite{Borsanyi:2020mff}. These gauge ensembles are at nearly physical pion and kaon mass. We plan to exploit up to six different lattice spacings in the range $[0.0640-0.1315]$ fm and consider $L=3$, $4$ and $6$ fm boxes for finite-size effect studies. Simulations are performed in the isospin limit where $m_u = m_d \equiv m_\ell$. In Table \ref{tab:ensembles} we summarize the two ensembles that are considered in this preliminary study for the $\pi^0$ and $\eta,\etap$ TFFs.
\begin{table}[h]
	\centering
	\begin{tabular}{p{1.6cm}|p{1cm}| p{1cm} | p{1.6cm}}
		&$\beta$& $a$[fm] & $L/a\times T/a$ \\
		\hline
		$\pi^0$ TFF & 4.0126 & 0.0640 & 96 $\times$ 144 \\
		$\eta,\etap$ TFF&3.7000& 0.1315 & 32 $\times$ 64 
	\end{tabular}
	\caption{Summary of two ensemble's parameters used for this preliminary result.}
	\label{tab:ensembles}
\end{table}
\section{Transition Form Factor on the Lattice}
\noindent The TFF of a pseudoscalar meson is defined by the matrix elements $M_{\mu\nu}$
\begin{align}
	M_{\mu\nu}(p,q_1) = i\int d^4x\, e^{iq_1\cdot x}\bra{\Omega}T\{J_\mu(x)J_\nu(0)\}\ket{P(\vec{p})}=\epsilon_{\mu\nu\alpha\beta} q_1^\alpha q_2^\beta \mathcal{F}_{P\gamma^*\gamma^*}(q_1^2,q_2^2)
\end{align}
where $q_1$ and $q_2$ are the photon $4$-momenta, $J_\mu$ is the hadronic component of the electromagnetic (EM) current and $\epsilon_{\mu\nu\alpha\beta}$ is the $4$-rank Levi-Civita tensor. These matrix elements $M_{\mu\nu}$ are related to a three-point correlation function $C_{\mu\nu}^{(3)}$ that is computed on the lattice \cite{Ji_2001, Gerardin:2016cqj}
\begin{equation}
	C_{\mu\nu}^{(3)}(\tau,t_{P}) =a^6 \sum_{\vec{x},\vec{z}} \langle J_\mu (\vec{z},t_i)J_\nu(\vec{0},t_f)P^\dagger(\vec{x},t_0)\rangle e^{i\vec{p}\cdot \vec{x}}e^{-i\vec{q}_1\cdot \vec{z}}
\end{equation}
where $t_{P} = min(t_f - t_0, t_i-t_0)$ is the minimal time separation between the pseudoscalar density and the vector currents and $\tau = t_i - t_f$ is the time-separation between the two EM currents. In the Euclidean:
	\begin{equation}
		M_{\mu\nu}^E = \frac{2E_{P}}{Z_{P}}\int_{-\infty}^{\infty}d\tau\, e^{\omega_1\tau}\tilde{A}_{\mu\nu}(\tau)
	\end{equation}
where 
\begin{equation}
	\tilde{A}_{\mu\nu}(\tau) \equiv \lim\limits_{t_{P}\to\infty} e^{E_{P}(t_f-t_0)} C_{\mu\nu}^{(3)}(\tau,t_{P}).
\end{equation}
$E_{P}$ is the energy of the pseudoscalar and $Z_{P}$ is the overlap factor of the meson with our choice of interpolating operator; they are extracted from two-point correlations functions (for details on the two-point function analysis see \cite{Verplanke:2021gat}). The momenta are $q_1 = (\omega_1,\vec{q}_1)$ and $q_2 = (E_{P} - \omega_1, \vec{p} - \vec{q}_1)$, where $\omega_1$ is a free parameter that determines what virtuality regime of the TFF is considered.

The correlation function $C_{\mu\nu}^{(3)}(\tau,t_{P})$ receives potential contributions from four Wick contractions, shown in Figure \ref{fig:wick}. From top to bottom we refer to them as PVV, P-VV, PV-V and P-V-V. In the case of $\pi^0$, our pseudoscalar interpolator is $P_{\pi^0} = \frac{1}{\sqrt{2}} \left(\overline{u}\gamma_5 u(x) - \overline{d}\gamma_5 d(x)\right)$. Since we work in the isospin limit, diagrams P-VV and P-V-V do not contribute. Furthermore, the remaining disconnected contribution, PV-V, has been shown to be small \cite{Gerardin:2019vio} and is not included in this preliminary analysis. For the the $\eta,\eta^{\prime}$ mesons the pseudoscalar interpolators take the form 
\begin{align*}
	P_{\eta_8}(x)= \frac{1}{\sqrt{6}}\left(\overline{u}\gamma_5 u(x) + \overline{d}\gamma_5 d(x) - 2\overline{s}\gamma_5 s(x)\right),\\
	P_{\eta_0}(x) = \frac{1}{\sqrt{3}} \left(\overline{u}\gamma_5 u(x) + \overline{d}\gamma_5 d(x) + \overline{s}\gamma_5 s(x)\right).
\end{align*}
As a consequence, all four possible Wick contractions contribute to $C_{\mu\nu}^{(3)}(\tau,t_{P})$. Especially P-VV has a large and noisy contribution, that spoils the signal quality. On top of that, the $\eta_8$ and $\eta_0$ mix to create the physical $\eta,\eta^{\prime}$ and this needs to be taken into account when computing the TFFs. 
\section{Pion Transition Form Factor}
\begin{figure}
	\centering
	\includegraphics[width=0.8\textwidth]{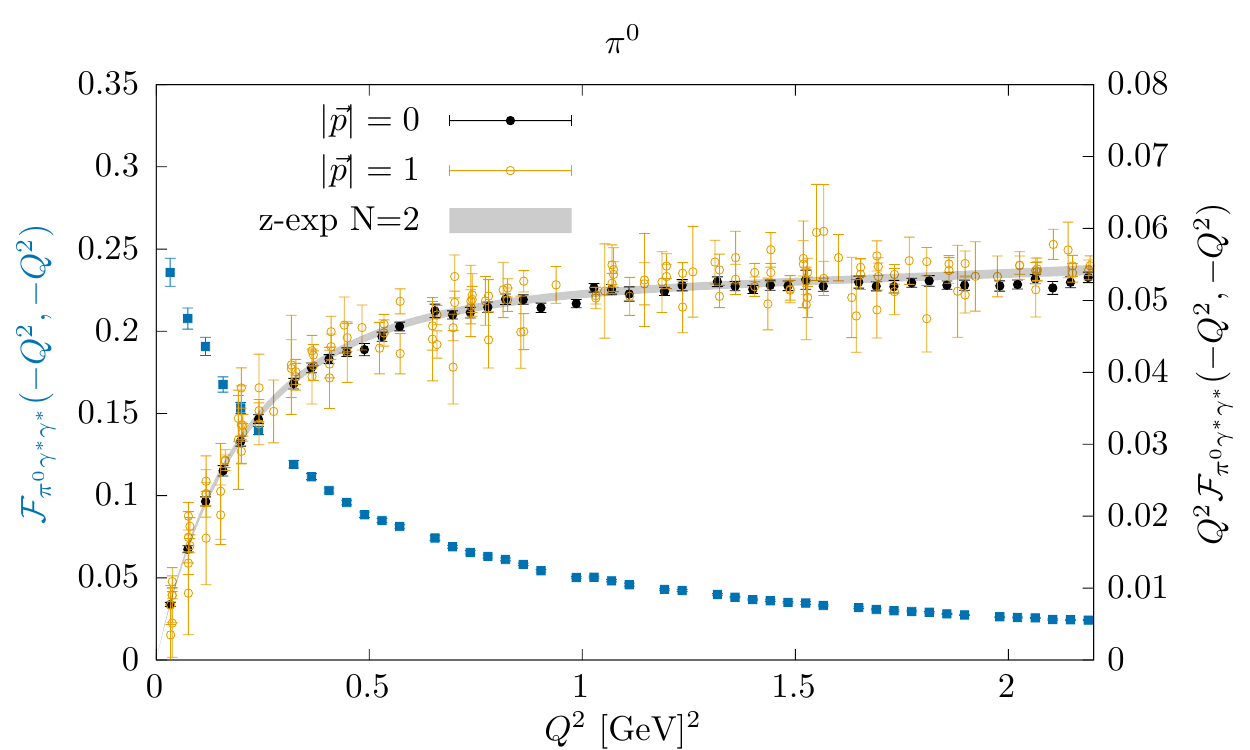}
	\caption{TFF in the doubly virtual regime. Right y-axis: $\vec{p} = \vec{0}$ frame (black filled circles) and $\vec{p} = \frac{2\pi}{L}(0,0,1)$ frame (yellow open circles); left y-axis: Frame $\vec{p} = \vec{0}$ (blue filled squares). Grey band indicates the result of a $z$-expansion fit with $N=2$ over all kinematical regimes. }
	\label{fig:pi_tff}
\end{figure}

\noindent We first consider the pion TFF, that is simpler to compute on the lattice than the $\eta,\etap$, and can be cross-checked with previous computations on the lattice \cite{Gerardin:2016cqj, Gerardin:2019vio}. In Figure \ref{fig:pi_tff} we plot our result for the TFF in the doubly virtual regime. First, we find a good agreement between the two reference frames of the pion at $\vec{p} = \vec{0}$ and $\vec{p} = \frac{2\pi}{L}(0,0,1)$. We also observe a plateau for $Q^2\mathcal{F}_{\pi^0\gamma^*\gamma^*}(-Q^2,-Q^2)$ at large $Q^2$ as predicted by the OPE at short distances \cite{nesterenko1983comparison, novikov1984use}. Additionally, looking at $\mathcal{F}_{\pi^0\gamma^*\gamma^*}(-Q^2,-Q^2)$ we see that the error of the TFF grows quickly with decreasing $Q^2$. This illustrates the challenge for a precise determination of $\amupole$, keeping in mind that the weight functions in Eq. (\ref{eq:amupole}) are peaked at low $Q^2$. 
To obtain a continuous description of the TFF in the whole kinematical range, that can used to evaluate Eq. (\ref{eq:amupole}), we fit out data using a modified $z$-expansion \cite{Gerardin:2016cqj}
\begin{align}
	P(Q_1^2,Q_2^2)\mathcal{F}_{\pi^0\gamma^*\gamma^*}(-Q_1^2,-Q_2^2) = \sum_{n,m=0}^{N}c_{nm}&\left(z_1^n +(-1)^{N+n}\frac{n}{N+1}z_1^{N+1}\right)\cdot\nonumber \\
	&\left(z_2^m+(-1)^{N+m}\frac{m}{N+1}z_2^{N+1}\right),
\end{align}
where $z_k$ are conformal variables
\begin{align*}
	z_k = \frac{\sqrt{t_c + Q_k^2} - \sqrt{t_c - t_0 \vphantom{Q_k^2}}}{\sqrt{t_c + Q_k^2} + \sqrt{t_c - t_0\vphantom{Q_k^2}}}, \quad k = 1,2,
\end{align*}
$c_{nm}$ are symmetric coefficients, $t_c = 4m_\pi^2$ maps the branch cut of TFF onto the unit circle $\left| z_k \right|= 1$ and $t_0$ is a free parameter (chosen such that the maximum value of $\left|z_k\right|$ is diminished in the given momentum range); $P(Q_1^2,Q_2^2)$ imposes short-distance constraints to aid the fit at large $Q^2$. The clear advantage of this expansion is that the fit is model-independent, the only systematic being the choice of $N$ in the sum.

Performing a $z$-expansion with $N=2$ gives a preliminary value $\amupipole = 63.3(2.9)\cdot 10^{-11}$, where the error is purely statistical. This value, obtained at a single lattice spacing, is comparable in magniture to the Mainz result \cite{Gerardin:2019vio} and with a competitive precision.

\section{Study of Finite-size Effects}
\noindent The pion TFF has been computed on $L=6$ fm boxes at each value of the lattice spacing as quoted in \cite{Borsanyi:2020mff}. This is possible since relatively few diagrams need to be computed here, and therefore the simulation cost does not become prohibitively expensive on these large boxes. In the case of the $\eta,\eta^{\prime}$ TFFs the noise/signal ratio increases rapidly due to large disconnected contributions. So it would be useful to use smaller volumes for this observable to be able to generate more statistics. To do so, however, we need to ensure that finite-size effects (FSE) do not play an important role for our observables. 

\begin{figure}[]
	\centering
	\subfloat[]{\includegraphics[width=0.5\textwidth]{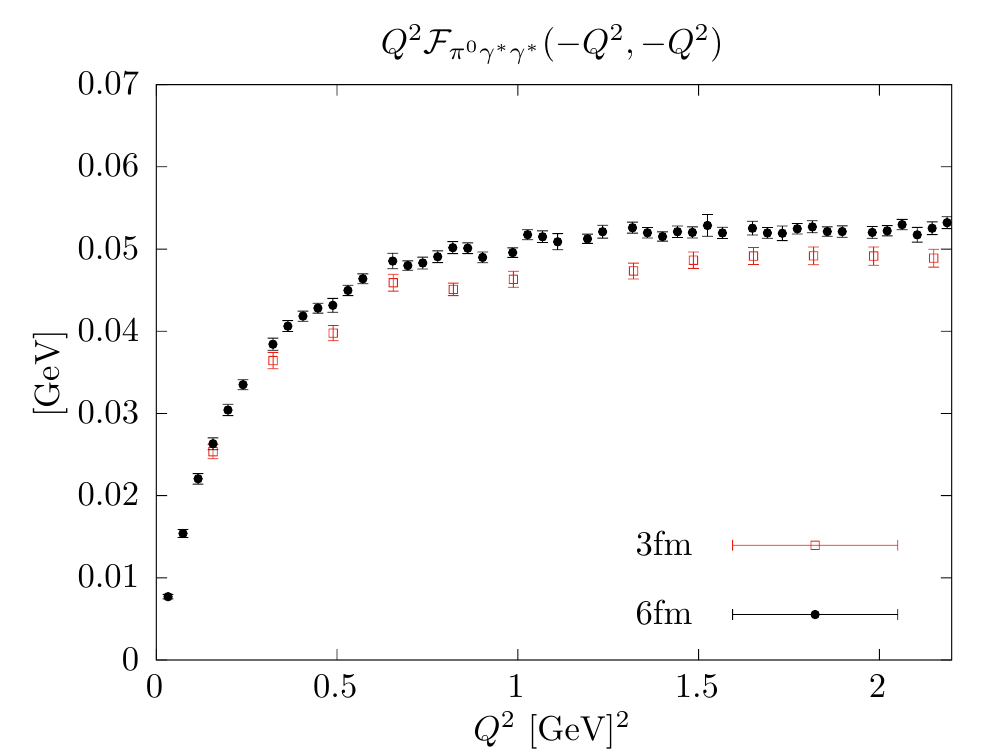}\label{fig:nofse}}
	\hfill
	\subfloat[]{\includegraphics[width=0.5\textwidth]{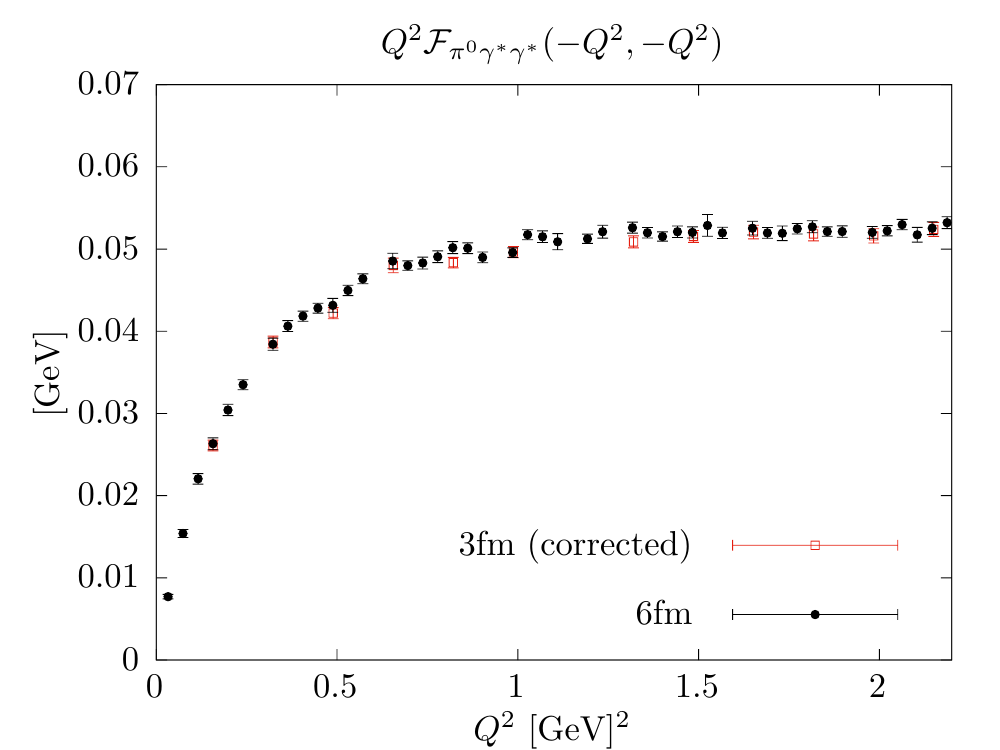}\label{fig:fse}}
	\caption{Pion TFF in the doubly virtual regime at two different volumes ($L =3,6$ fm). Before (left) and after (right) correcting for backward propagating pion.}
	\label{fig:fsefse}
\end{figure}

To test this possibility, we study the FSE for the $\pi^0$, for which a high precision has been achieved. In Figure \ref{fig:nofse} we plot the TFF in the doubly virtual regime for $L = 6$ fm and $ L = 3$ fm boxes. Here, we have not yet taken into account possible significant effects of backward propagating pions as noted and demonstrated in \cite{Gerardin:2016cqj}. In fact, when one corrects for this effect\footnote{Note that this correction depends exponentially on the energy of the pseudoscalar meson. Since we work with taste-singlet pions, the correction is the largest at finest lattice spacing, where the taste-singlet pion is lighter and volume effects are more likely to be apparent.}, the data for the two different box sizes agree well as can be seen in Figure \ref{fig:fse}. Since the $\eta,\etap$ are even heavier mesons, we decide to also use $3$fm and $4$fm boxes to compute the $\eta,\eta^{\prime}$ TFFs.

\section{$\eta,\eta^{\prime}$ Transition Form Factors}
\noindent As mentioned before, in the case of the $\eta,\etap$ TFFs, we have four contributing diagrams and the mixing between the unphysical $\eta_8$ and $\eta_0$ to create the $\eta,\eta^{\prime}$ states. Moreover, the noisy P-VV contribution is large and of opposite sign as compared to the PVV contribution, which complicates a precise determination of the TFFs. The use of smaller volumes allows us to  generate a lot of statistics and improve our signal/noise ratio. Further, as presented in \cite{verplanke_muong2}, we apply analysis techniques, alongside the brute force increase of statistics, to improve our signal. 

\begin{figure}[]
	\centering
	\subfloat[]{\includegraphics[width=0.4\textwidth]{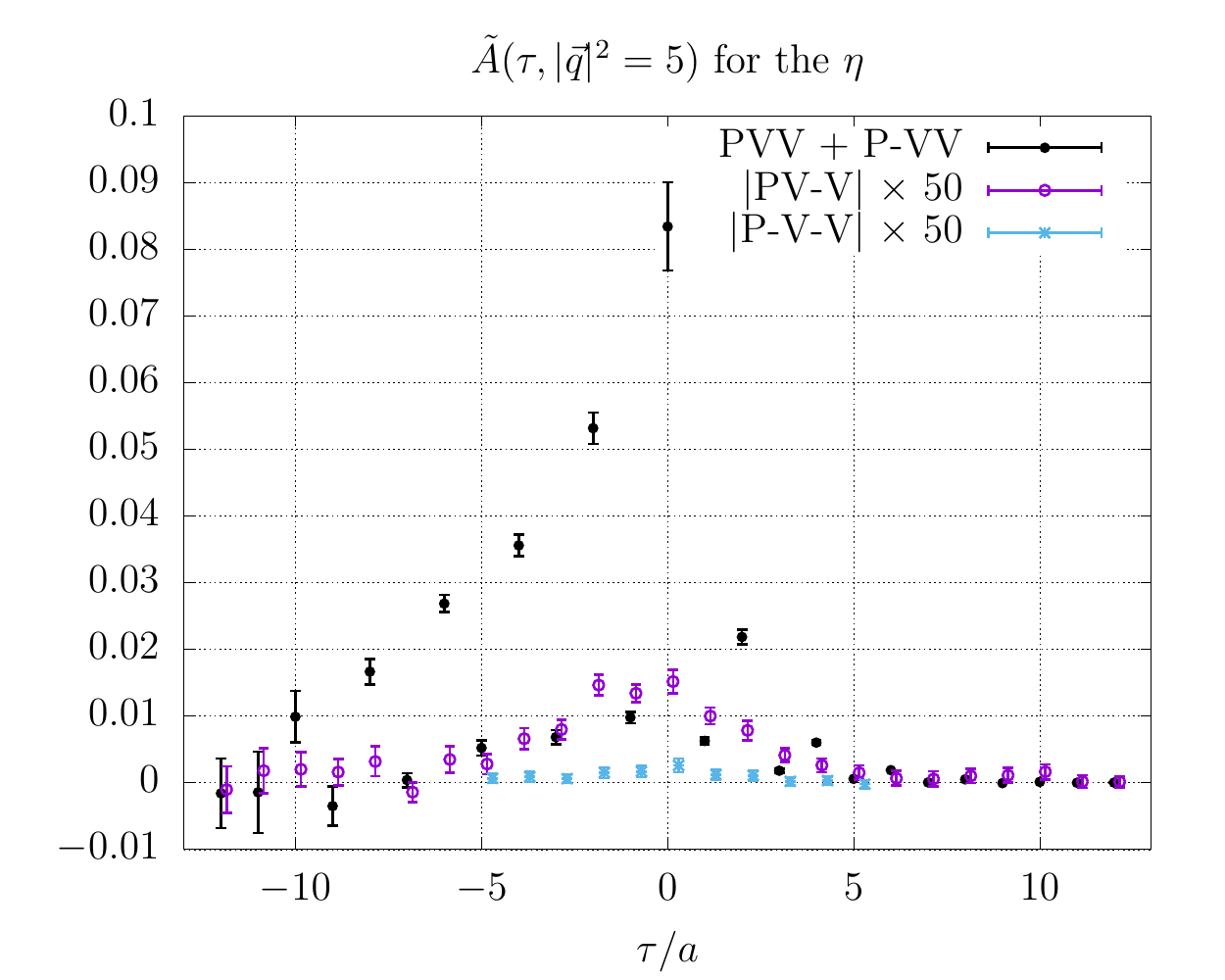}\label{fig:int_eta}}
	\hfill
	\subfloat[]{\includegraphics[width=0.35\textwidth]{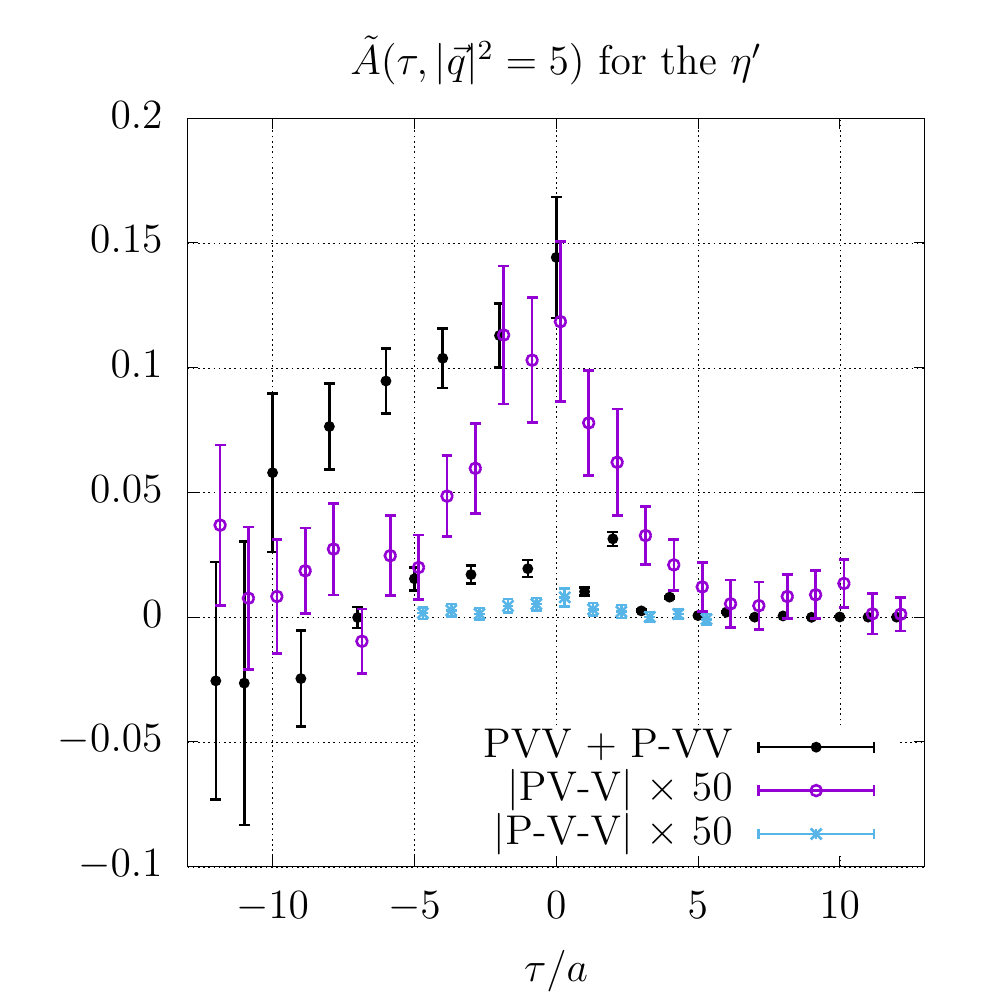}\label{fig:int_etap}}
	\hfill
	\subfloat[]{\includegraphics[width=0.2\textwidth]{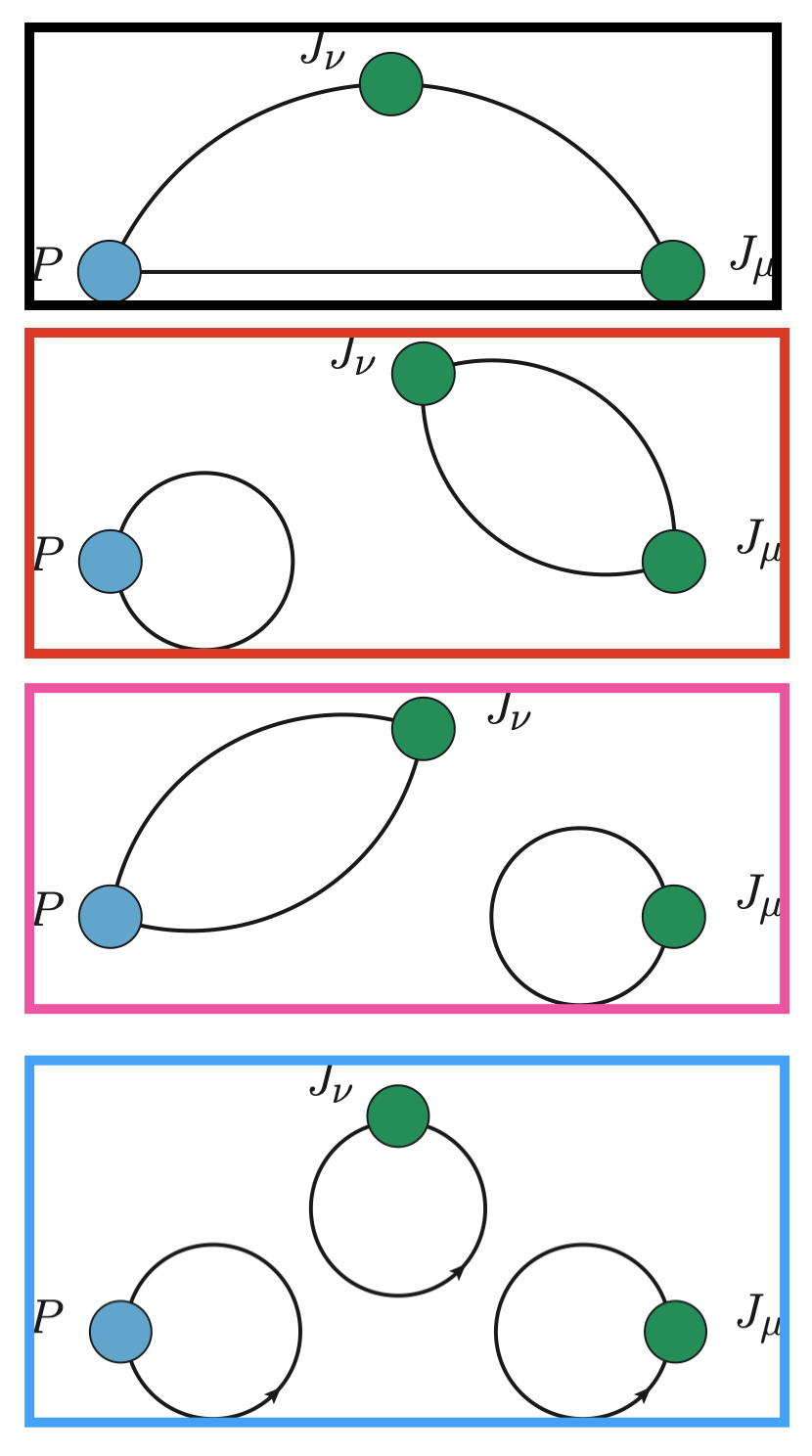}\label{fig:wick}}
	\caption{Left: Integrand of the $\eta$ TFF. Points have been shifted in the horizontal direction for clarity. Middle: Integrand of the $\eta^{\prime}$ TFF. Right: Different possible contractions contributing to the three-point function $C_{\mu\nu}^{(3)}(\tau,t_{P})$.}
	\label{fig:integr}
\end{figure}

\begin{figure}
	\centering
	\includegraphics[width=\textwidth]{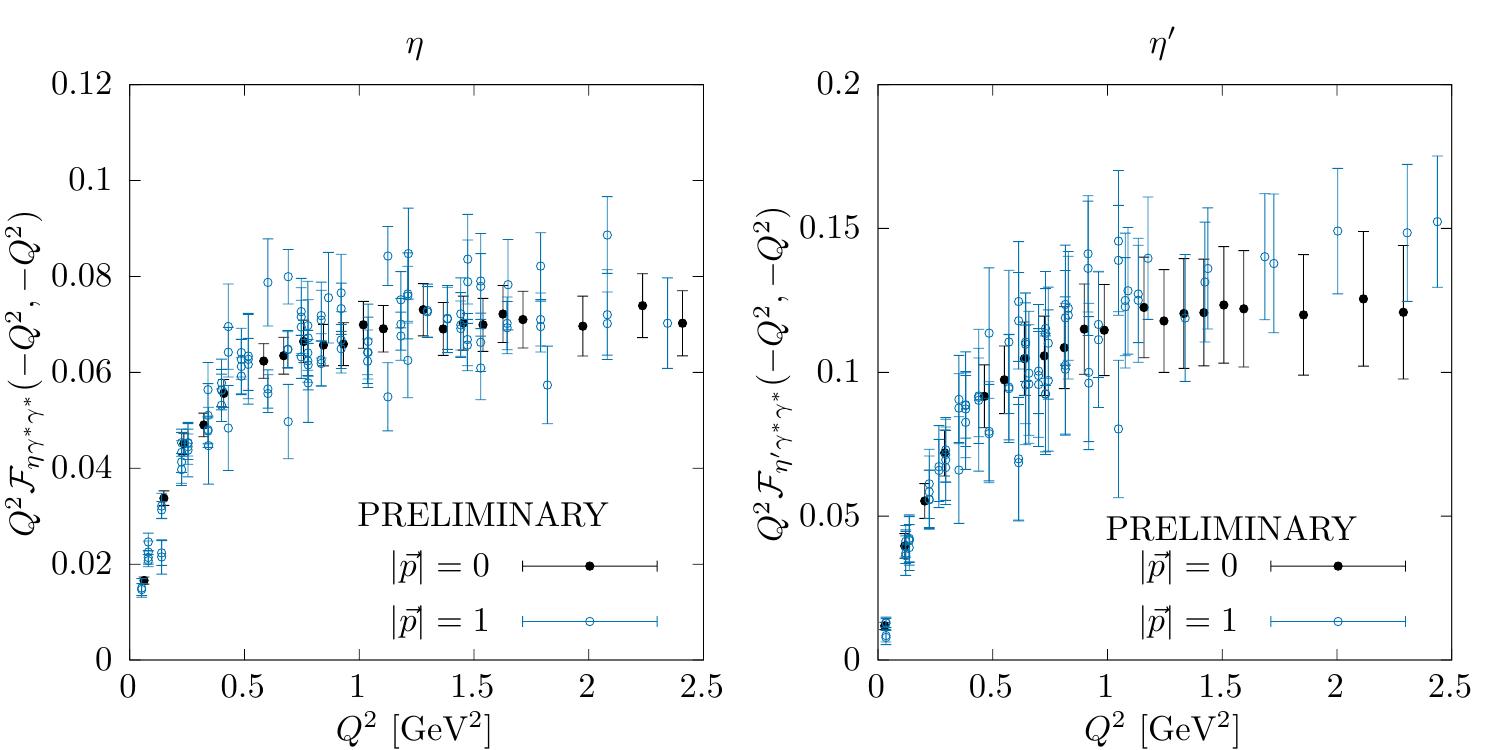}
	\caption{TFF of the $\eta$ (left), $\eta^{\prime}$ (right) in the doubly virtual regime. Black filled circles and blue open circles indicate respectively $\vec{p}=\vec{0}$ and $\vec{p} = \frac{2\pi}{L}(0,0,1)$.}
	\label{fig:eta_doubly}
\end{figure}

\begin{figure}
	\centering
	\includegraphics[width=\textwidth]{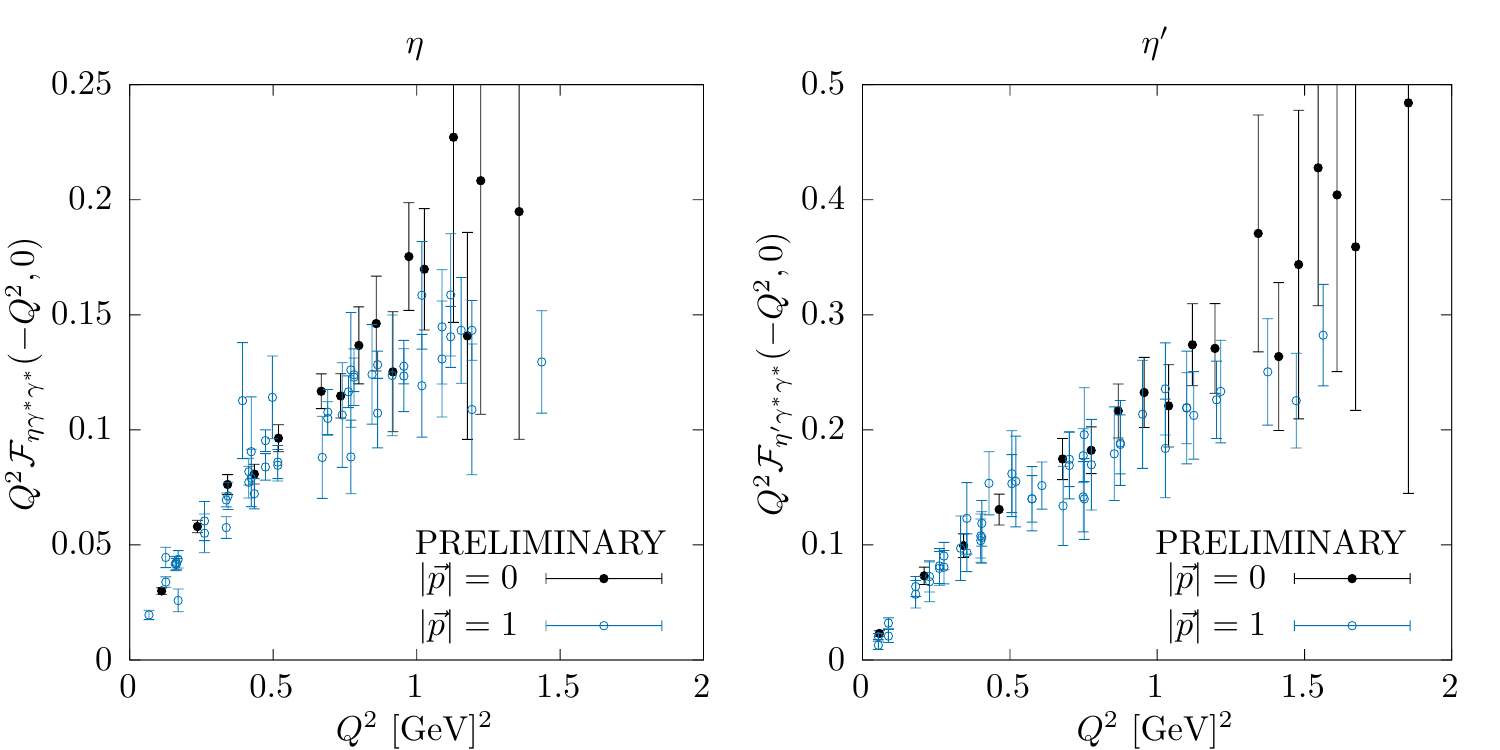}
	\caption{TFF of the $\eta$ (left), $\eta^{\prime}$ (right) in the singly virtual regime. Black filled circles and blue open circles indicate respectively $\vec{p}=\vec{0}$ and $\vec{p} = \frac{2\pi}{L}(0,0,1)$.}
	\label{fig:eta_singly}
\end{figure}

The result for the integrands of the $\eta,\etap$ TFFs are shown in Figure \ref{fig:integr}. The leading contribution is the sum of the PVV and P-VV diagrams; the PV-V and P-V-V are comparably smaller. A preliminary calculation of the TFFs yields results presented in Figures \ref{fig:eta_doubly} and \ref{fig:eta_singly}. There is a good agreement between the two reference frames of the $\eta,\etap$ with $\vec{p} = \vec{0}$ \& $\vec{p} = \frac{2\pi}{L}(0,0,1)$. Errors are larger than for the $\pi^0$ precisely because of the difficulties mentioned before. We also see that the signal looks promising in the two different kinematical frames, particularly important for the singly virtual regime since it enters directly into the formula for \amupole. A preliminary $z$-expansion fit on the data leads to $\amuetapole = 28(5)\cdot 10^{-11}$ and $\amuetappole = 30(10)\cdot 10^{-11}$, where the error is purely statistical. The results are relatively large compared to other estimates \cite{Masjuan:2017tvw, Eichmann_2019, Raya_2020}, but we stress that values are computed at our coarsest lattice spacing, and a dedicated continuum extrapolation still needs to be performed.
\section{Acknowledgements}
\noindent This publication received funding from the Excellence Initiative of Aix-Marseille University - A*MIDEX, a French “Investissements d’Avenir” programme, AMX-18-ACE-005 and from the French National Research Agency under the contract ANR-20-CE31-0016. This work was granted access to the HPC resources of TGCC under the allocation 2022-A0120511504 made by GENCI. Center de Calcul Intensif d'Aix-Marseille is acknowledged for granting access to its high performance computing resources.

\newpage
\bibliographystyle{JHEP}
\begingroup
\fontsize{10pt}{3pt}\selectfont
\bibliography{bibliography}
\endgroup

\end{document}